\documentclass[final,5p,times,onecolumn]{elsarticle}
\usepackage{amssymb}
\usepackage[linesnumbered,lined,ruled]{algorithm2e}
\usepackage{amsmath}
\usepackage{subfigure}
\usepackage{amsthm}
\usepackage[usenames,dvipsnames]{color}
\usepackage{threeparttable}
\usepackage{textcomp}
\usepackage[T1]{fontenc}
\usepackage{booktabs}
\usepackage{multirow}
\usepackage{setspace}
\usepackage{stfloats}
\usepackage{fancyhdr}
\usepackage{geometry}
\usepackage{lineno,hyperref}
\modulolinenumbers[5]
\biboptions{sort&compress}
\journal{Signal Processing}
\begin{document}
	\newcommand{\be}{\begin{equation}}
	\newcommand{\ee}{\end{equation}}
	\newcommand{\br}{{\mbox{\boldmath{$r$}}}}
	\newcommand{\bp}{{\mbox{\boldmath{$p$}}}}
	\newcommand{\bpi}{\mbox{\boldmath{ $\pi $}}}
	\newcommand{\bn}{{\mbox{\boldmath{$n$}}}}
	\newcommand{\balfa}{{\mbox{\boldmath{$\alpha$}}}}
	\newcommand{\ba}{\mbox{\boldmath{$a $}}}
	\newcommand{\bta}{\mbox{\boldmath{$\beta $}}}
	\newcommand{\bg}{\mbox{\boldmath{$g $}}}
	\newcommand{\bPsi}{\mbox{\boldmath{$\Psi $}}}
	\newcommand{\bsigma}{\mbox{\boldmath{ $\Sigma $}}}
	\newcommand{\bGamma}{{\bf \Gamma }}
	\newcommand{\bA}{{\bf A }}
	\newcommand{\bP}{{\bf P }}
	\newcommand{\bX}{{\bf X }}
	\newcommand{\bI}{{\bf I }}
	\newcommand{\bR}{{\bf R }}
	\newcommand{\bZ}{{\bf Z }}
	\newcommand{\bz}{{\bf z }}
	\newcommand{\bx}{{\mathbf{x}}}
	\newcommand{\bM}{{\bf M}}
	\newcommand{\bU}{{\bf U}}
	\newcommand{\bD}{{\bf D}}
	\newcommand{\bJ}{{\bf J}}
	\newcommand{\bH}{{\bf H}}
	\newcommand{\bK}{{\bf K}}
	\newcommand{\bm}{{\bf m}}
	\newcommand{\bN}{{\bf N}}
	\newcommand{\bC}{{\bf C}}
	\newcommand{\bL}{{\bf L}}
	\newcommand{\bF}{{\bf F}}
	\newcommand{\bv}{{\bf v}}
	\newcommand{\bSigma}{{\bf \Sigma}}
	\newcommand{\bS}{{\bf S}}
	\newcommand{\bs}{{\bf s}}
	\newcommand{\bO}{{\bf O}}
	\newcommand{\bQ}{{\bf Q}}
	\newcommand{\btr}{{\mbox{\boldmath{$tr$}}}}
	\newcommand{\bNSCM}{{\bf NSCM}}
	\newcommand{\barg}{{\bf arg}}
	\newcommand{\bmax}{{\bf max}}
	\newcommand{\test}{\mbox{$
			\begin{array}{c}
			\stackrel{ \stackrel{\textstyle H_1}{\textstyle >} } { \stackrel{\textstyle <}{\textstyle H_0} }
			\end{array}
			$}}
	\newcommand{\tabincell}[2]{\begin{tabular}{@{}#1@{}}#2\end{tabular}}
	\newtheorem{Def}{Definition}
	\newtheorem{Pro}{Proposition}
	\newtheorem{Exa}{Example}
	\newtheorem{Rem}{Remark}
	\begin{frontmatter}
		
		\title{Robust Poisson Multi-Bernoulli Mixture Filter with Unknown Detection Probability}
		
		\author{Guchong Li}
		\author{Lingjiang Kong\corref{mycorrespondingauthor}}
		\cortext[mycorrespondingauthor]{Corresponding author}
		\ead{lingjiang.kong@gmail.com}
		\author{Tao Zhou}
		\author{Tuanwei Tian}
		\author{Wei Yi}
		
		\address{School of Information and Communication Engineering, University of Electronic Science and Technology of China, Chengdu, 611731, China}
		\fntext[myfootnote]{This work was supported in part by the National Natural Science Foundation of China under Grant 61771110, in part by the Chang Jiang Scholars Program, in part by the 111 Project No. B17008.}
%

\begin{abstract}
This paper proposes a robust \emph{Poisson multi-Bernoulli mixture} (R-PMBM) filter immune to the unknown detection probablity. In a majority of multi-object scenarios, the prior knowledge of detection probability is usually uncertain, which is often estimated offline from the training data. In such cases, online filtering is always unfeasible or unrealistic, otherwise, significate parameter mismatch will result in biased estimates (e.g., state and cardinality of objects). As a consequence, the ability of adaptively estimating the detection probability for a sensor is essential in practice. Based on the analysis, we detail how the detection probability can be estimated accompanied with the state estimates. Besides, the closed-form solutions to the proposed method are derived by means of approximating the intensity of Poisson random finite set (RFS) to a Beta-Gaussian mixture (B-GM) form and density of Bernoulli RFS to a single Beta-Gaussian form. Simulation results demonstrate the effectiveness and robustness of the proposed method.
\end{abstract}
\end{frontmatter}

\section{Introduction}
Multi-object tracking (MOT) has been an increasingly hot topic both in military and civilian areas in the last few years. The aim of MOT is to jointly estimate the state and cardinality of objects synchronously from the monitored scenario. So far, MOT has been widely adopted in many fields, such as enviromental monitoring, battlefield surveillance and distributed sensor network \cite{BarShalom1988,Blackman1999,BarShalom2001,Mahler2014,Asyn2019,KVVJ2019}. 

However, a common difficulty for MOT is the association problem between objects and observations. Amongst currently studied algorithms, Joint Probabilistic Data Association (JPDA) \cite{BarShalom1988}, Multiple Hypotheses Tracking (MHT) \cite{Blackman1999}, and Random Finite Set (RFS) \cite{Mahler2000,MahlerPHD} are the main solutions to MOT. In particular, RFS approaches receive a high degree of attention due to the effective solutions to the association problem. Under the FInite Set STatistics (FISST) framework, some filters based on the RFS theory are developed mainly comprised by two 
types: unlabeled and labeled filters. The former (unlabeled RFS-based filters) mainly consists of Probability Hypothesis Density (PHD) \cite{MahlerPHD}, Cardinality-PHD (CPHD) \cite{MahlerCPHD} and multi-Bernoulli (MB) \cite{Vo2009,Vo2010}, as well as the recently developed Poisson MB mixture (PMBM) \cite{Williams2012,Williams2016,Angel2018} filters, while the latter (labeled RFS-based filters) includes labeled MB (LMB) \cite{Reuter2014}, generalized LMB (GLMB) \cite{Vo2013,Vo2014}, labeled MBM (LMBM)\cite{Mahler2016}, and marginalized $\delta$-GLMB (M$\delta$-GLMB) \cite{Fantacci2016} filters. Moreover, all of the mentioned filters can be effectively implemented by resorting to either Gaussian mixture (GM) \cite{Vo2006,VVC2007,VVC2009,AYKLJ2019} or sequential Monte Carlo (SMC) \cite{Vo2005,Sidenbladh2003,Krofreiter2016Fusion} technologies. Comparing to the SMC method, the GM method can provide a closed-form solution. 

Comparing to the other unlabeled RFS-based filters (PHD, CPHD, and MB), a unique and important characteristic of the PMBM filter \cite{Williams2016,Angel2018} is the conjugacy property, which means that the posterior distribution has the same functional form as the prior. It is also proofed that the labeled RFS-based GLMB \cite{Vo2013} and LMBM \cite{Mahler2016} filters are conjugated. The reason why conjugacy property is important is that it allows the posterior to be written in terms of some single-object predictions and updates, which provides a convenient computation method compared with the direct caculation of multi-object predictions and updates. Further, the PMBM filter also shows advantages in low-detection scenarios \cite{Williams2016,Xia2017,Angel2018,Guchong2019}. As a consequence, the PMBM filter has received a lot of attention since it was proposed. and has been increasingly adopted in many applications \cite{Karl2018,Samuel2018,Maryam2017,FLGW2019}.

It is worth noting that when processing the real world data there is a significant source of certainly, detection model, in addition to the dynamic model, i.e., the detection probability in radar tracking is always related to the detection distance, weather and so on, making it difficult to model accurately. In general, the detection model is usually assumed to be known by the offline estimate from the training data in most algorithms. In such cases, online filtering process for the filters mentioned above is not feasible, otherwise, significant mismatch in detection model  will cause erroneous estimates of both state and cardinality of objects. In order to make the filters more adaptable to the environment, Mahler et al. have proposed a robust CPHD (R-CPHD) filter by online estimating the unknown detection probability \cite{Mahler2010,Mahler2011} in which object state is augmented with a parameter of detection model and the augmented state model is propagated and estimated along with the R-CPHD recursion. The proposed R-CPHD filter in \cite{Mahler2011} has been applied to track cell microscopy data with unknown background parameters \cite{Rezatofighi2015}. Afterwards, the similar idea has also been successfully applied to the MB filter \cite{VVHM2013} and the labeled RFS-based filter \cite{PVVK2018}. Both of them show the effectiveness of the proposed strategy in \cite{Mahler2011}. Recently, another method by exploiting the Inverse Gamma Gaussian mixture (IGGM) distribution to implement the PHD/CPHD filters is also proposed in \cite{LWKSP2018}, and a GLMB-based method for multistatic Doppler radar with unknown detection probability is studied in \cite{CH2019}. To the best of our knowledge, the research on the PMBM filter with unknown detection probability hasn't been realized yet. 

Considering the attractive characteristics of the PMBM filter such as conjugacy property and low detection tolerance, we explore a robust PMBM (R-PMBM) recursion subject to an unknown detection probability jointly estimating the state of object and detection probability. The main contributions of the paper are described as follows:
\begin{enumerate}
	\item In Section 3, \emph{we propose an effective R-PMBM recursion immune to the unknown detection model for.} Firstly, the state of object is coupled with a variable representing the detection probability so that the standard PMBM filtering process is evolved into a R-PMBM filter which can jointly estimate the state of objects and detecion probability. Next, the expressions of the proposed R-PMBM filter recursion are given.
	\item In Section 4, \emph{we present a computationally feasible implementation of the proposed R-PMBM filter by resorting to a Beta function to depict the detection probability.} Except for the state of objects, the detection model is also needed to consider during the proposed robust PMBM filtering process. To model the detection probability, Beta distribution is selected, where detection probabilty can be easily extracted by seeking the expectation of the Beta distribution. Moreover, the Gaussian distribution is still used to model the kinematic state, which is the same as the standard PMBM filter. As a consequence, the closed-form Beta-Gaussian mixture is constructed as a menas of implementation.	
	\item In Section 5, \emph{four simulation experiments are given to verify the effectiveness and robustness of the proposed method.} In order to better verify the low detection tolerance of the proposed method, two cases with different detection probabilities are considered. Further, for each case, two groups of simulation experiments between R-CPHD and R-PMBM filters are provided to compare the covariance of observation noise and clutter rate respectively.
\end{enumerate}

The outline of the rest of the paper is as follows. Section 2 introduces the background knowledge, and Section 3 describes the PMBM filter with unknown detection probability, and its corresponding detailed implementation is provided in Section 4. Simulation results are provided in Section 5, and conclusions are drawn in Section 6.

\section{Background}

\subsection{Notations}
In this paper, lower case letters (e.g., $x$ and $z$) denote state and observation of single-object while upper case letters (e.g., $X$ and $Z$) denote state and observation of multi-object, respectively. Suppose there are $N$ objects and $M$ observations at time $k$, and then the \emph{multi-object state} and \emph{multi-object observation} are modelled as RFSs given by 
\begin{align}
{X}_k &= \left\{ {x_{k,1}, \cdots ,x_{{k,N}}} \right\} \subset {{\cal X}}, \\
{Z}_k &= \left\{ {z_{k,1}, \cdots ,z_{{k,M}}} \right\} \subset {{\cal Z}},
\end{align}
where $\cal X$ and $\cal Z$ denote the state space and observation space respectively. Each single-object state $x_{k,i} = {[x_{k,p}^i,x_{k,v}^i]^{\top}}$ comprises the position $x_{k,p}^i$ and velocity $x_{k,v}^i$, where $`\top'$ denotes the transpose.

For a set $X$ and a function $f(x)$, 
\begin{align}
[f(\cdot)]^X = \prod\limits_{x\in X}f(x).
\end{align}
The cardinality of a set $X$ is denoted $\left| X \right|$. $\uplus$ is denoted as the disjoint set uinon. Given $X^u\uplus X^d = X$, $X^u$ and $X^d$ satisfy $X^u\cup X^d=X$ and $X^u\cap X^d=\emptyset$.
\subsection{Multi-object Bayes Filter}
Given \emph{multi-object transition function} $f_{k|k-1}(\cdot|\cdot)$ and the multi-object state $f_{k-1}(X|Z_{1:k-1})$ atC time $k-1$, where $Z_{1:{k-1}}$ is an array of finite sets of observations received up to time $k-1$ and denoted as $Z_{1:k-1}=\left(Z_1,\cdots
,Z_{k-1}\right)$, the \emph{multi-object prediction} to time $k$ can be given according to the Chapman-Kolmogorov equation
\begin{align}
{f_{k|k - 1}}({X_k|Z_{1:k-1}}) &= \int {{f_{k|k - 1}}({X_k}|\xi){f_{k - 1}}(\xi|Z_{1:k-1})\delta \xi}.
\end{align}

When a new set of observations $Z_k$ is received at time $k$, which is modeled as a \emph{multi-object observation likelihood} $g_k(Z_k|X_k)$, the \emph{multi-object update} at time $k$ is given based on multi-object Bayes rule
\begin{align}
{f_k}({X_k|Z_{1:k}}) &= \frac{{{g_k}({Z_k}|{X_k}){f_{k|k - 1}}({X_k|Z_{1:k-1}})}}{{\int {{g_k}({Z_k}|\xi){f_{k|k - 1}}(\xi|Z_{1:k-1})\delta \xi} }},
\end{align}
where the involved integral is the set integral \cite{Mahler2014} which is defined by 
\begin{equation}
\nonumber
\int {f(X)\delta X}  = \sum\limits_{n = 0}^\infty  {\frac{1}{{n!}}} \int_{{X^n}} {f(\{ {x_1}, \cdots ,{x_n}\} )d{x_1} \cdots d{x_n}}.
\end{equation}

Further, for the convenience of representation, we leave out the condition on the observation set $Z_{1:k}$ and abbreviate $f_k(X_k|Z_{1:k})$ as $f_k(X)$.
\subsection{PMBM RFS}
Before the PMBM density is introduced, two neccessary definitions are present to help understand the PMBM RFS.
\begin{Def}
	Undetected objects are those objects that exist at the current time but have never been detected and denoted by $X_k^u$.
\end{Def}
\begin{Def}
	A new observation may be a new object for the first detection and can also correspond to another previously detected object or clutter. Considering that it may exist or not, we refer to it as a potentially detected object denoted by $X_k^d$.
\end{Def}
Conditioned on the observation set $Z_{1:k}$, the multi-object state RFS $X_k$ is modeled as the union of independent RFS $X_k^u$ (undetected objects) and $X_k^d$ (potentially detected objects), respectively. Hence, the posterior density of the PMBM RFS can be denoted by the FISST convolution as
\begin{equation}
f_k\left( X \right) = \sum\limits_{Y \subseteq X} {{f_k^{{\mathop{\rm p}\nolimits}}}(Y){f_k^{{{\mathop{\rm mbm}\nolimits}}}}(X-Y)}.
\end{equation}
$f_k^{{\mathop{\rm p}\nolimits}}(\cdot)$ is a Poisson density given by
\begin{align}
\nonumber
f_k^{{\mathop{\rm p}\nolimits}}(X) &= {e^\lambda }\prod\limits_{i = 1}^n {\lambda f_k({x_i})}\\
\label{PPP2}
&= {e^{ - \int {\mu_k (x)dx} }}{\left[ {\mu_k ( \cdot )} \right]^{X}},
\end{align}
where $\mu_k(x)=\lambda f_k(x)$ is the intensity function and $\lambda$ the Poisson rate as well as $f_k(x)$ a probability density function (pdf) of a single object.
Moreover, $f_k^{{{\mathop{\rm mbm}\nolimits}}}(\cdot)$ is a MBM density given by
\begin{equation}
\label{MBM}
{f_k^{{\mathop{\rm mbm}\nolimits}}}(X) \propto \sum\limits_{j \in {\mathbb I}} {\sum\limits_{{X_{1}} \uplus  \cdots  \uplus {X_{n}} = X} {\prod\limits_{i = 1}^n {{\omega_{j,i}}{f_{j,i}}({X_i})} } },
\end{equation}
where $`\propto'$ denotes the proportional symbol. It can be seen that the MBM RFS is the normalized and weighted sum of multi-object densities of MBs, which is parameterized by\\ ${\left\{ {{w_{j,i}}, {{\left\{ {{r_{j,i}},{f_{j,i}}(x)} \right\}}_{i \in {{\mathbb I}^j}}}} \right\}_{j \in {\mathbb I}}}$, where $\mathbb I$ is the index set of the MBs (also called global hypothesis set). Particularly, the MBM RFS degenerates into the MB RFS 
\begin{equation}
\label{MB}
{f_k^{{\mathop{\rm mb}\nolimits}}}(X) \propto  {\sum\limits_{{X_1} \uplus  \cdots  \uplus {X_n} = X} {\prod\limits_{i = 1}^n {{\omega_{j,i}}{f_{j,i}}({X_i})} } }
\end{equation}
when there is only one global hypothesis with $\left|\mathbb I \right|=1$.
\subsection{PMBM Recursion}
Here, a review of the recursive processes (prediction and update) of the PMBM filter is given.
\subsubsection{Prediction Process}
Poisson density $f_{k-1}^{\mathop{\rm p}\nolimits}(\cdot)$ and the MBM density $f_{k-1}^{{\mathop{\rm mbm}\nolimits}}(\cdot)$ are predicted separately.
\begin{enumerate}[(a)]
	\item Suppose the intensity function of Poisson density at time $k-1$ is $\mu_{k-1}(x)$, and then the predicted intensity at time $k$ is
	\begin{equation}\label{Pre_poisson}
	\begin{split}
	\mu_{k|k-1} (x) = {\gamma _k}(x)+ \int {f_{k|k-1}(x|\xi){p_{S,k}}(\xi){\mu_{k-1}}(\xi)d\xi}
	\end{split}
	\end{equation}
	where ${\gamma _k}(x)$ is the intensity of birth model at time $k$ and $f_{k|k-1}(x|\xi)$ and $p_{S,k}(\cdot)$ denote the state transition function of single object and survival probability, respectively.
	\item Given the $i$-th object in the $j$-th global hypothesis at time $k-1$ with ${\omega_{k - 1}^{j,i},r_{k - 1}^{j,i},p_{k - 1}^{j,i}({x})}$, and then the prediction process of MBM components is given by
	\begin{align}
	\label{Pre_bernoulli_w}
	\omega_{k|k-1}^{j,i} &= \omega_{k-1}^{j,i}, \\
	\label{Pre_bernoulli_r}
	{r_{k|k-1}^{j,i}} &= r_{k-1}^{j,i}\int {p_{k-1}^{j,i}(\xi){p_{S,k}}(\xi)d\xi}, \\
	\label{Pre_bernoulli_p}
	{p_{k|k-1}^{j,i}}(x) &\propto \int {f_{k|k-1}(x|\xi){p_{S,k}}(\xi)p_{k-1}^{j,i}(\xi)d\xi},
	\end{align}
	where $\omega_{k|k-1}^{j,i},r_{k|k-1}^{j,i},p_{k|k-1}^{j,i}(x)$ denote the predicted hypothesis weight, existence probability, and pdf of the $i$-th Bernoulli component in the $j$-th global hypothesis, respectively.
\end{enumerate}

\subsubsection{Update Process}
The update process mainly consists of the following four parts:
\begin{itemize}
	\item update for undetected objects;
	\item update for potential objects detected for the first time;
	\item misdetection for previous potentially detected objects;
	\item update for previous potentially detected objects using received observation set.
\end{itemize}
The specific expressions are given in (a)-(d), respectively.
\begin{enumerate}[(a)]
	\item Update for undetected objects: \begin{equation}\label{undetected_Poisson}
	{\mu_k}(x) = { {(1 - {p_{D,k}(x)})\mu_{k|k-1} (x)}},
	\end{equation}
	where $p_{D,k}(\cdot)$ is the detection probability.
	\item Update for potential objects detected for the first time: 
	\begin{align}
	\label{potential_MB_r}
	{r_k^{{\mathop{\rm p}\nolimits}}}(z) &= {{e_k(z)} \mathord{\left/  \right.} {{\rho_k ^{{\mathop{\rm p}\nolimits}}}(z)}},\\
	\label{potential_MB_p}
	{p_k^{{\mathop{\rm p}\nolimits}}}(x|z) &= {p_{D,k}(x)}g_k(z|x)\mu_{k|k-1} (x)/e_k(z),
	\end{align}
	and
	\begin{align}
	\label{weight_potential_object}
	{\rho_k ^{{\mathop{\rm p}\nolimits}}}(z) &= e_k(z) + c(z),\\
	e_k(z) &= \int {g_k(z|\xi){p_{D,k}(\xi)}\mu_{k|k-1} (\xi)d\xi}
	\end{align}
	where $\rho_k^{{\mathop{\rm p}\nolimits}}(z)$ in (\ref{weight_potential_object}) is the hypothesis weight of the potential object related to the observation $z$.
	\item Misdetection for previous potentially detected objects:
	\begin{align}
	\label{miss1}
	{\omega_k^{j,i}}(\emptyset) &= {\omega_{k|k-1}^{j,i}}(1 - {p_{D,k}(x)}{r_{k|k-1}^{j,i}}),\\
	\label{miss2}
	{{r}_k^{j,i}}(\emptyset) &= {r_{k|k-1}^{j,i}}(1 - {p_{D,k}(x)})/(1 - p_{D,k}(x){r_{k|k-1}^{j,i}}),\\
	\label{miss3}
	p_k^{j,i}(\emptyset) &= p_{k|k-1}^{j,i}(x).
	\end{align}
	\item Update for previous potentially detected objects using received observation set: 
	\begin{align}
	{{\omega}_k^{j,i}}(z) &= {\omega_{k|k-1}^{j,i}}{r_{k|k-1}^{j,i}} \int {{p_{D,k}(\xi)}g_k(z|\xi){p_{k|k-1}^{j,i}}(\xi)d\xi}, \\
	r_k^{j,i}(z) &= 1,\\
	\label{weight_meas}
	{{p}_k^{j,i}}(x,z) &\propto {p_{D,k}(x)}g_k(z|x){p_{k|k-1}^{j,i}}(x).
	\end{align}
\end{enumerate}

It can be seen that the update process is also separate where the undetected objects are just preserved by multiplying the weight with a misdetection probability shown in (\ref{undetected_Poisson}) and the potential detected targets are updated consisting of three parts (parts (b)-(d)).

After updating single-object density, another factor to be considered is to generate global hypothesis set. Essentially, the global hypothesis should undergo all possible data association based on all single-object hypotheses. To settle the computation bottleneck, \emph{Murty's algorithm} \cite{Murty1968} is
selected as an achievable skill in which a cost matrix is constructed by the calculated weight in (\ref{potential_MB_p}), (\ref{miss2}) and (\ref{weight_meas}) (see (\ref{cost_matrix}) in Section III-B for the construction method). The detailed implementation steps can be referred to \cite{Angel2018}.

\section{The proposed R-PMBM Filter}
\begin{figure}[htbp]
	\centering
	\includegraphics[width=3.1in]{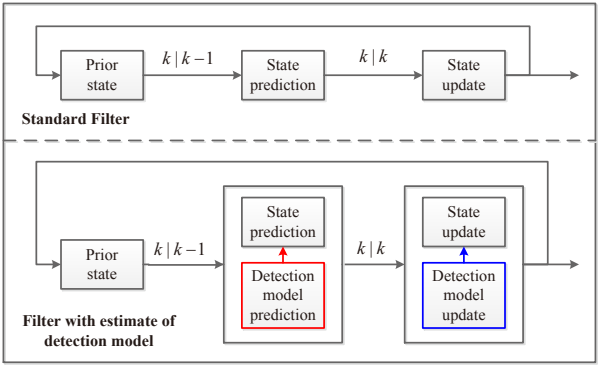}
	\caption{The comparison between the standard filter and proposed filter with estimate of detection model.}
	\label{fig:shiyitu}
\end{figure}

In this section, the joint estimates of state of objects and detection probability are provided. Firstly, the basic construction method is introduced by augmenting a variable denoting the unknown detection probability to each state of object. Hereafter, the proposed R-PMBM recursion for the augmented state model is derived. Moreover, a comparison of filtering framework between the standard filter and the proposed filter is described in Fig. \ref{fig:shiyitu}. 
\subsection{Augmented State Model}
Following the approach in \cite{Mahler2011}, a variable $a\in[0,1]$ representing the detection probability is augmented to the state of object $x$,
\begin{align}
\hat{x} = (x,a).
\end{align}
The integral of the augmented state $\hat{x}$ is adjusted into a double integral,
\begin{align}
\int {f(\hat x)d\hat x}  = \int {\int_0^1 {f(x,a)dadx} }.
\end{align} 

Meanwhile, the state transition and observation models are the same as the conventional case, except that we focus on the augmented state model, which are given by
\begin{eqnarray}
\nonumber
{f_{k|k - 1}}(\hat x|\hat \zeta ) &=& {f_{k|k - 1}}(x,a|\zeta ,\alpha )\\
&=& {f_{k|k - 1}}(x|\zeta ){f_{k|k - 1}}(a|\alpha ),\\
{g_k}(z|\hat x) &=& {g_k}(z|x,a) = {g_k}(z|x), \\
{p_{S,k}}(\hat x) &=& {p_{S,k}}(x,a) = {p_{S,k}}(x),\\
{p_{D,k}}(\hat x) &=& {p_{D,k}}(x,a) = a.
\end{eqnarray}
Furthermore, the birth model with the augmented model is denoted as an intensity function $\lambda _k^b(x,a)$.
\subsection{Recursion}
The derivation of the R-PMBM filter recursion for the augmented state model featuring the unknown detection probability is straightforward by substituting the sugmented state model into the standard PMBM filter recursion. Next, the direct consequences of derivion are given by Propositions 1 and 2. 
\begin{Pro}
	If at time $k-1$, the intensity of Poisson RFS $\mu_{k-1}(\hat{x})$ and MB RFS with $\{\omega_{k-1}^{j,i},r_{k-1}^{j,i},p_{k-1}^{j,i}(\hat{x})\}$ are given, which denote the undetected objects and potential objects respectively, then the predicted intensity of Poisson process and density of MBM process can be given by
	\begin{enumerate}[(a)]
		\item Poisson Process: 
		\begin{align}
		\nonumber
		&{\mu _{k|k - 1}}(x,a) = {\gamma _k}(x,a)\\
		\label{PP1}
		&+ \int {\int_0^1 {{f_{k|k - 1}}(x|\zeta ){f_{k|k - 1}}(a|\alpha ){p_{S,k}(\zeta)}} } {\mu _{k - 1}}(\zeta,\alpha )d\alpha d\zeta.
		\end{align}
		\item MBM Process:
		\begin{align}
		\label{PB1}
		\omega_{k|k - 1}^{j,i} &= \omega_{k - 1}^{j,i},\\
		\label{PB2}
		r_{k|k - 1}^{j,i} &= r_{k - 1}^{j,i}\int {\int_0^1 {{p_{S,k}}(\zeta)p_{k - 1}^{j,i}(\zeta,\alpha )d\alpha d\zeta} },\\
		\label{PB3}
		p_{k|k - 1}^{j,i}(x,a) 		
		&\propto \int {\int_0^1 {{p_{S,k}}(\zeta){f_{k|k - 1}}(x|\zeta ){f_{k|k - 1}}(a|\alpha )p_{k - 1}^{j,i}(\zeta,\alpha )d\alpha d\zeta} }. 
		\end{align}
	\end{enumerate}
\end{Pro}
\begin{Pro}
	If at time $k$, the predicted R-PMBM filter with parameters $\{\mu_{k|k-1}(\hat{x}),\omega_{k|k-1}^{j,i},r_{k|k-1}^{j,i},p_{k|k-1}^{j,i}(\hat{x})\}$ is given, then for a given observation set $Z_k$, the updated intensity of Poisson process and density of MBM process can be given from four aspects.
	\begin{enumerate}[(a)]
		\item Update for undetected objects: 
		\begin{equation}
		\label{update_PPP}
		{\mu _{k|k}}(x,a) = \left( {1 - a} \right){\mu_{k|k-1}}(x,a).
		\end{equation}
		\item Update for potential objects for the first time:
		\begin{align}
		\label{update_b1}
		r_k^{{\mathop{\rm p}\nolimits}}(z) &= {e_k}(z)/\rho _k^{{\mathop{\rm p}\nolimits}}(z), \\
		\label{update_b2}
		p_k^{{\mathop{\rm p}\nolimits}}(x,a|z) &= a{g_k}(z|x)\mu _{k|k - 1}(x,a)/{e_k}(z),
		\end{align}
		where 
		\begin{align}
		\label{weight_potrntial}
		\rho _k^{{\mathop{\rm p}\nolimits}}(z) &= {e_k}(z) + c(z), \\
		\label{update_b3}
		{e_k}(z) &= \int {\int_0^1 {\alpha{g_k}(z|\zeta) \mu _{k|k - 1}(\zeta,\alpha )d\alpha d\zeta} }.
		\end{align}
		\item Misdetection for potentially detected objects:
		\begin{align}
		\label{weight_empty}
		\omega_{k}^{j,i}(\emptyset )&= \omega_{k|k - 1}^{j,i} \times (1-r_{k|k-1}^{j,i}+\varsigma^{j,i}), \\
		\label{update_c1}
		r_{k}^{j,i}(\emptyset )
		&= \frac{{r_{k|k - 1}^{j,i}\int {\int_0^1 {p_{k|k - 1}^{j,i}(\zeta,\alpha )(1 - \alpha )d\alpha d\zeta} } }}{{1 - r_{k|k - 1}^{j,i} + \varsigma^{j,i} }}, \\
		\label{update_c2}
		p_{k}^{j,i}(\emptyset ,a) &= \frac{{p_{k|k - 1}^{j,i}(x,a)(1 - a)}}{{\int {\int_0^1 {p_{k|k - 1}^{j,i}(\zeta,\alpha )(1 - \alpha )d\alpha d\zeta} } }},\\
		\label{update_c3}
		\varsigma^{j,i} &= r_{k|k - 1}^{j,i}\int {\int_0^1 {p_{k|k - 1}^{j,i}(\zeta,\alpha )(1 - \alpha )d\alpha d\zeta} }. 
		\end{align}
		\item Update for previous potentially detected objects using received observation set:
		\begin{align}
		\nonumber
		\omega_{k}^{j,i}(z) =& \omega_{k|k - 1}^{j,i}r_{k|k - 1}^{j,i}\\
		\label{update_d2}
		&\times \int {\int_0^1 {\alpha {g_k}(z|\zeta)p_{k|k - 1}^{j,i}(\zeta,\alpha )d\alpha d\zeta} },\\
		\label{update_d3}
		r_{k}^{j,i}(z) =& 1, \\
		p_{k}^{j,i}(x,a|z) =& \frac{{a{g_k}(z|x)p_{k|k - 1}^{j,i}(x,a)}}{{\int {\int_0^1 {\alpha {g_k}(z|\zeta)p_{k|k - 1}^{j,i}(\zeta,\alpha )d\alpha d\zeta} } }}.
		\end{align}
	\end{enumerate}
\end{Pro}

The global hypotheses are returned by selecting some new single-object hypotheses for the next recursion.
In order to avoid all possible data hypotheses for each previous global hypothesis, we still adopt the construction strategy based upon the \emph{Murty's algorithm} \cite{Murty1968}. Assume there are $n_o$ old tracks in the global hypothesis $j$ and $M$ observations $\{z_1,\cdots,z_M\}$, which indicates that there are $M$ potential detected objects. Then the cost matrix at time $k$ can be formed as follows.
\begin{equation}\label{cost_matrix}
C_j = -\text{ln} \left[ {\begin{array}{*{20}{c}}
	{\digamma_p}&{\digamma_o}
	\end{array}}  \right]_{M\times(M+n_o)}
\end{equation}
with
\begin{align}
\nonumber
\digamma_p &=  
\left[ {\begin{array}{*{20}{c}}
	{\nu_{j}^{1,1}}&{\nu_{j}^{1,2}}& \cdots &\nu_j^{1,n_o}\\
	\vdots & \vdots & \ddots & \vdots \\
	{\nu_{j}^{M,1}}&{\nu_{j}^{M,2}}& \cdots &\nu_j^{M,n_o}
	\end{array}} \right]_{M\times n_o}\\
\nonumber
\digamma_o &= 
\left[ {\begin{array}{*{20}{c}}
	{\nu_p^{1,1}}& \cdots & 0\\
	\vdots & \ddots & \vdots \\
	0 & \cdots &{\nu_p^{M,M}}
	\end{array}} \right]_{M\times M}
\end{align}
where $\nu_p^{m,m}$ ($m \in \{1,\cdots,M\}$) denotes the weight of the $m$-th potential detected object given by (\ref{weight_potrntial}) and  $\nu_{j}^{m,n}$ is the weight corresponding to the $m$-th observation updated by the $n$-th old track in the $j$-th global hypothesis, which is denoted as
\begin{equation}\label{weight_update}
\nu_{j}^{m,n} = {\omega_{k|k-1}^{j,i}}{\rho_{k}^{j,i}}(z)/{\rho_{k}^{j,i}}(\emptyset)
\end{equation}
with ${\rho_{k}^{j,i}}(z)$ given by
\begin{equation}
\label{cost_matrix2}
\rho _k^{j,i}(z) = r_{k|k-1}^{j,i} \int{ \int {\alpha{g_k}(z|\zeta )p_{k|k - 1}^{j,i}(\zeta,\alpha)d\alpha }d\zeta},
\end{equation}
and ${\rho _k^{j,i}}(\emptyset)$ given by (\ref{weight_empty}). 

It is worth noting the weights of global hypotheses need to be normalized once the construction process about global hypothesis set is finished. A summative description of the proposed algorithm steps is provided in Algorithm 1. Further, the global hypothesis with the largest weight is selected during the state estimate.

\begin{Rem}
	The proposed method has a similar but sightlying higher complexity compared to the standard PMBM recursion. This is because the filter framework is unchanged but an additional variable/funcction needs to be propagated corresponding to many more components to be maintained.
\end{Rem}

\begin{algorithm}[htb]
	\DontPrintSemicolon
	\caption{Description of the proposed R-PMBM filter.}
	\LinesNumbered
	\small\underline{\sc {INPUT}}: $\mu_{k-1}(\hat x)$, $\{\omega_{k-1}^{j,i},r_{k-1}^{j,i},p_{k-1}^{j,i}(\hat x)\}$, $\gamma_k(\hat{x})$.\;
	\textbf{-Perform Prediction:}\\
	\text{--Poisson Process:} {$\mu_{k-1}(\hat{x}) \rightarrow \mu_{k|k-1}(\hat{x})$}  $\hspace{18.5mm} \rhd (\ref{PP1})$\\
	\text{--MB Process:}\\ {$\{\omega_{k-1}^{j,i},r_{k-1}^{j,i},p_{k-1}^{j,i}(\hat{x})\} \rightarrow \{\omega_{k|k-1}^{j,i},r_{k|k-1}^{j,i},p_{k|k-1}^{j,i}(\hat{x})\}$}\\ $\hspace{60.5mm} \rhd(\ref{PB1})$-$(\ref{PB3})$\\
	\textbf{-Perform Update:}\\
	\text{--Update for undetected objects:} {$\mu_{k|k-1}(\hat{x}) \rightarrow \mu_{k}(\hat{x})$}  $\hspace{3mm}\rhd (\ref{update_PPP})$\\
	\text{--Update for potential objects detected for the first time:} $\{r_k^{{\mathop{\rm p}\nolimits}}(z),{p_k^{{\mathop{\rm p}\nolimits}}}(\hat{x}|z),{\rho_k^{{\mathop{\rm p}\nolimits}}}(z)\}$ $\hspace{32mm} \rhd (\ref{update_b1})$-$(\ref{update_b3})$\\
	\text{--Misdetection for previous potentially detected objects:}
	$\{\omega_k^{j,i}(\emptyset),r_k^{j,i}(\emptyset),p_k^{j,i}(\emptyset,a)\}$ $\hspace{27mm}\rhd (\ref{weight_empty})$-$(\ref{update_c3})$\\
	\text{--Update for previous potentially detected objects:}      $\{\omega_k^{j,i}(z),r_k^{j,i}(z),p_k^{j,i}(\hat{x},z)\}$ $\hspace{28mm}\rhd (\ref{update_d2})$-$(\ref{update_d3})$\\
	\textbf{-Construct Global hypotheses:}\\
	\text{--Form cost matrix:} $C_j$ $\hspace{30.5mm}\rhd (\ref{cost_matrix})$-$(\ref{cost_matrix2})$\\
	\text{-Run Murty's algorithm;}\\
	\text{-Normalize the weight of global hypotheses set.}\\
	\small\underline{\sc {OUPUT}}: $\mu_{k}(\hat x)$, $\{\omega_{k}^{j,i},r_{k}^{j,i},p_{k}^{j,i}(\hat x)\}$.\;
\end{algorithm}

\section{Beta-Gaussian Mixture Implementation}
In this section, a closed-form implementation for the proposed R-PMBM recursion immune to the unknown detection probability is derived based on the Beta-Gaussian mixture. The Gaussian distribution is used to model the state of object same as the standard PMBM filter while the Beta function is used to model the detection probability. Before the implementation is given, the definition and some properties about the Beta distribution are first provided as follows.
\begin{Def}
	For $0 \le a \le 1$, the shape parameters $s>1$, $t>1$, are a power function of variable $a$ and of its reflection ($1-a$) as follows.
	\begin{align}
	\beta \left( {a;s,t} \right) = \frac{{{a^{s - 1}}{{(1 - a)}^{t - 1}}}}{{\int_0^1 {{\alpha ^{s - 1}}{{(1 - \alpha )}^{t - 1}}d\alpha } }} = \frac{{{a^{s - 1}}{{(1 - a)}^{t - 1}}}}{{B\left( {s,t} \right)}}
	\end{align}
	with mean ${\mu _\beta } = {\textstyle{s \over {s + t}}}$ and covariance $\sigma _\beta ^2 = {\textstyle{{st} \over {{{(s + t)}^2}(s + t + 1)}}}$. $\beta(a;s,t)$ and ${B\left( {s,t} \right)}$ are named Beta distribution and Beta function respectively.
\end{Def}
For the Beta distribution $\beta \left( {a;s,t} \right)$, some of its properties are summarized which will be used in the following deriviation.
\begin{align}
\label{pro1}
(1 - a)\beta (a;s,t) &= {{{B(s,t + 1)} \over {B(s,t)}}}\beta (a;s,t + 1), \\
\label{pro2}
a\beta (a;s,t) &= {{{B(s + 1,t)} \over {B(s,t)}}}\beta (a;s + 1,t), \\
\label{pro3}
\frac{s}{{s + t}} &= \frac{{B(s + 1,t)}}{{B(s,t)}}, \\
\label{pro4}
\frac{t}{{s + t}} &= \frac{{B(s,t + 1)}}{{B(s,t)}}.
\end{align}
Moreover, the prediction of the Beta distribution satisfies
\begin{align}
\label{pro5}
\beta \left( {{a_{\rm{ + }}};{s_{\rm{ + }}},{t_{\rm{ + }}}} \right) = \int {\beta \left( {a;s,t} \right){f_ + }({a_ + }|a)da}
\end{align}
with
\begin{align}
\nonumber
{s_{\rm{ + }}} &= \left( {{{{{\mu _{\beta , + }}(1 - {\mu _{\beta , + }})} \over {\sigma _{\beta , + }^2}}}} \right){\mu _{\beta , + }},\\
\nonumber
{t_{\rm{ + }}} &= \left( {{{{{\mu _{\beta , + }}(1 - {\mu _{\beta , + }})} \over {\sigma _{\beta , + }^2}}}} \right)(1 - {\mu _{\beta , + }}), \\
\nonumber
\sigma_{\beta,+}^2 &= k_\beta \sigma_\beta^2, k_\beta \ge 1.
\end{align}

For the considered standard linear Gaussian model, some assumptions are given as follows.
\begin{itemize}
	\item Each object follows a linear Gaussian dynamical model, i.e,
	\begin{align}
	{f_{k|k - 1}}(x|\zeta ) &= {\cal N}(x;{F_{k - 1}}\zeta ,{Q_{k - 1}}), \\
	{g_k}(z|x) &= {{\cal N}}(z;{H_k}x,{R_k}),
	\end{align}
	where $F_{k-1}$ and $Q_{k-1}$ denote the state transition matrix and process noise covariance, and $H_{k}$ and $R_{k}$ are the observation matrix and observation noise covariance, respectively.
	\item The survival probability for each object is state independent, i.e,
	\begin{align}
	p_{S,k}(x) = p_{S,k}.
	\end{align}
	\item The intensity of newborn model is a Beta-Gaussian mixture form
	\begin{align}
	{\gamma _k}(x,a) = \sum\limits_{i = 1}^{{J_k^{\gamma}}} {\eta _{r,k}^i\beta (a;s_{r,k}^i,t_{r,k}^i){{\cal N}}(x;m_{r,k}^i,P_{r,k}^i)},
	\end{align} 
	where $J_k^{\gamma}$, $\eta_{r,k}^i$, $s_{r,k}^i$, $t_{r,k}^i$, $m_{r,k}^i$, $P_{r,k}^i$, $i=1,\cdots,J_k^{\gamma}$ are given model parameters.
\end{itemize}

Then, the analytic solution to the R-PMBM filter with unknown detection probability can be represented in Propositions 3 and 4.
\begin{Pro}
	If at time $k-1$, the intensity of Poisson process  $\mu_{k-1}(\hat{x})$ is a Beta-Gaussian mixture form
	\begin{align}
	{\mu _{k - 1}}(x,a) = \sum\limits_{i = 1}^{{J_{k - 1}^u}} {\eta _{k - 1}^{i,u}\beta (a;s_{k - 1}^{i,u},t_{k - 1}^{i,u}){{\cal N}}(x;m_{k - 1}^{i,u},P_{k - 1}^{i,u})},
	\end{align}
	and the denisty of $i$-th Bernoulli component in the $j$-th hypothesis is a single Beta-Gaussian form
	\begin{align}
	p_{k - 1}^{j,i}(x,a) = {\beta (a;s_{k - 1}^{j,i},t_{k - 1}^{j,i}){{\cal N}}(x;m_{k - 1}^{j,i},P_{k - 1}^{j,i})},
	\end{align}
	then, the predicted intensity of Poisson process and density of MBM process are given by
	\begin{enumerate}[(a)]
		\item Poisson Process:
		\begin{align}
		\nonumber
		&{\mu _{k|k - 1}}(x,a) = {\gamma _k}(x,a)\\
		&+ \sum\limits_{i = 1}^{J_{k - 1}^u } {\eta _{k|k - 1}^{i,u}\beta (a;s_{k|k - 1}^{i,u},t_{k|k - 1}^{i,u}){{\cal N}}(x;m_{k|k - 1}^{i,u},P_{k|k - 1}^{i,u})}
		\end{align}
		where
		\begin{align}
		\nonumber
		s_{k|k - 1}^{i,u} =& \left( {\frac{{\mu _{\beta ,k|k - 1}^{i,u}(1 - \mu _{\beta ,k|k - 1}^{i,u})}}{{{{[\sigma _{\beta ,k|k - 1}^{i,u}]}^2}}} - 1} \right)\mu _{\beta ,k|k - 1}^{i,u},\\
		\nonumber
		t_{k|k - 1}^{i,u} =& \left( {\frac{{\mu _{\beta ,k|k - 1}^{i,u}(1 - \mu _{\beta ,k|k - 1}^{i,u})}}{{{{[\sigma _{\beta ,k|k - 1}^{i,u}]}^2}}} - 1} \right) (1 - \mu _{\beta ,k|k - 1}^{i,u}), \\
		\nonumber
		\eta_{k|k-1}^{i,u} =& p_{S,k}\eta_{k-1}^{i,u}, \\
		\nonumber
		m_{k|k - 1}^{i,u} =& {F_{k - 1}}m_{k - 1}^{i,u}, \\
		\nonumber
		P_{k|k - 1}^{i,u} =& {Q_{k - 1}} + {F_{k - 1}}P_{k - 1}^{i,u}F_{k - 1}^{\top}, \\
		\nonumber
		\mu _{\beta ,k|k - 1}^{i,u} =& \mu _{\beta ,k - 1}^{i,u} = {{{s_{k - 1}^{i,u}} \over {s_{k - 1}^{i,u} + t_{k - 1}^{i,u}}}},\\
		\nonumber
		{[\sigma _{\beta ,k|k - 1}^{i,u}]^2} =& {k_\beta }{[\sigma _{\beta ,k - 1}^{i,u}]^2}, \\
		\nonumber
		=& {{{s_{k - 1}^{i,u}t_{k - 1}^{i,u}} \over {{{(s_{k - 1}^{i,u} + t_{k - 1}^{i,u})}^2}(s_{k - 1}^{i,u} + t_{k - 1}^{i,u} + 1)}}}.
		\end{align}
		\item MBM Process:
		\begin{align}
		\omega_{k|k - 1}^{j,i} &= \omega_{k - 1}^{j,i}, \\
		r_{k|k - 1}^{j,i} &= {p_{S,k}}r_{k - 1}^{j,i}, \\
		p_{k|k - 1}^{j,i}(x,a) &= {\beta (a;s_{k|k - 1}^{j,i},t_{k|k - 1}^{j,i}){{\cal N}}(x;m_{k|k - 1}^{j,i},P_{k|k - 1}^{j,i})},
		\end{align}
		where 
		\begin{align}
		\nonumber
		s_{k|k - 1}^{j,i} =& \left( {\frac{{\mu _{\beta ,k|k - 1}^{j,i}(1 - \mu _{\beta ,k|k - 1}^{j,i})}}{{{{[\sigma _{\beta ,k|k - 1}^{j,i}]}^2}}} - 1} \right)\mu _{\beta ,k|k - 1}^{j,i}, \\
		\nonumber
		t_{k|k - 1}^{j,i} =& \left( {\frac{{\mu _{\beta ,k|k - 1}^{j,i}(1 - \mu _{\beta ,k|k - 1}^{j,i})}}{{{{[\sigma _{\beta ,k|k - 1}^{j,i}]}^2}}} - 1} \right) (1 - \mu _{\beta ,k|k - 1}^{j,i}), \\	
		\nonumber
		m_{k|k - 1}^{j,i} =& {F_{k - 1}}m_{k - 1}^{j,i}, \\
		\nonumber
		P_{k|k - 1}^{j,i} =& {Q_{k - 1}} + {F_{k - 1}}P_{k - 1}^{j,i}F_{k - 1}^{\top}, \\
		\nonumber
		\mu _{\beta ,k|k - 1}^{j,i} =& \mu _{\beta ,k - 1}^{j,i} = {{{s_{k - 1}^{j,i}} \over {s_{k - 1}^{j,i} + t_{k - 1}^{j,i}}}}, \\
		\nonumber
		{[\sigma _{\beta ,k|k - 1}^{j,i}]^2} =& {k_\beta }{[\sigma _{\beta ,k - 1}^{j,i}]^2}\\ 
		\nonumber
		=& {{{s_{k - 1}^{j,i}t_{k - 1}^{j,i}} \over {{{(s_{k - 1}^{j,i} + t_{k - 1}^{j,i})}^2}(s_{k - 1}^{j,i} + t_{k - 1}^{j,i} + 1)}}}.
		\end{align}
	\end{enumerate}
\end{Pro}
\begin{Rem}
	The proof is straightforward by substituting the Beta-Gaussian mixture form into the prediction equations in Proposition 1. The resultant expressions are also the Beta-Gaussian mixture form where the intensity of Poisson density is a Beta-Gaussian mixture form and the density of Bernoulli component is a single Beta-Gaussian form. The prediction of Gaussian distribution is the same as prediction in the standard GM-PMBM filter while that of Beta distribution is based on the property (\ref{pro5}) of the Beta distribution.
\end{Rem}

\begin{Pro}
	If at time $k$, the predicted intensity of Poisson density $\mu_{k|k-1}(x,a)$ is given by the following Beta-Gaussian mixture form
	\begin{align}
	\mu_{k|k-1}(x,a) 
	&= \sum\limits_{i = 1}^{J_{k|k - 1 }^u} {\eta _{k|k - 1}^{i,u}\beta (a;s_{k|k - 1}^{i,u},t_{k|k - 1}^{i,u}){{\cal N}}(x;m_{k|k - 1}^{i,u},P_{k|k - 1}^{i,u})},
	\end{align}	
	where $J_{k|k-1}^{u} = J_{k-1}^{u} + |\gamma_k|$, and the predicted density of $i$-th Bernoulli component in the $j$-th hypothesis is given by a Beta-Gaussian form
	\begin{align}
	p_{k|k - 1}^{j,i}(x,a) = {\beta (a;s_{k|k - 1}^{j,i},t_{k|k - 1}^{j,i}){{\cal N}}(x;m_{k|k - 1}^{j,i},P_{k|k - 1}^{j,i})},
	\end{align}
	then, given an observation set $Z_k$, the update of Poisson process and MBM process is given from four following parts.
	\begin{enumerate}[(a)]
		\item Update for undetected objects: 
		\begin{align}
		{\mu _{k}}(x,a) = \sum\limits_{i = 1}^{J_{k|k-1}^u } {\eta _{k,1}^{i,u}\beta (a;s_{k,1}^{i,u},t_{k,1}^{i,u}){{\cal N}}(x;m_{k,1}^{i,u},P_{k,1}^{i,u})},
		\end{align}
		where
		\begin{align}
		\nonumber
		\eta _{k,1}^{i,u } &= \eta _{k|k - 1}^{i,u }{\textstyle{{B(s_{k|k - 1}^{i,u },t_{k|k - 1}^{i,u } + 1)} \over {B(s_{k|k - 1}^{i,u },t_{k|k - 1}^{i,u })}}},\\
		\nonumber
		s_{k,1}^{i,u } &= s_{k|k - 1}^{i,u }, \\
		\nonumber
		t_{k,1}^{i,u } &= t_{k|k - 1}^{i,u } + 1, \\
		\nonumber
		m_{k,1}^{i,u } &= m_{k|k - 1}^{i,u },\\
		\nonumber
		P_{k,1}^{i,u } &= P_{k|k - 1}^{i,u }.
		\end{align}
		\item Update for potential objects for the first time:
		\begin{align}
		r_k^{{\mathop{\rm p}\nolimits}}(z) =& {e_k}(z)/\rho _k^{{\mathop{\rm p}\nolimits}}(z), \\
		p_k^{{\mathop{\rm p}\nolimits}}(x,a|z) =&
		\nonumber
		\frac{1}{{{e_k}(z)}}\sum\limits_{i = 1}^{J_{k|k - 1}^u } \eta _{k|k - 1}^{i,u}{\textstyle{{B(s_{k|k - 1}^{i,u} + 1,t_{k|k - 1}^{i,u})} \over {B(s_{k|k - 1}^{i,u},t_{k|k - 1}^i)}}} \\
		&\times \beta (a;s_{k,2}^{i,u} ,t_{k,2}^{i,u})q_{k,2}(z) {{\cal N}}(x;{m_{k,2}^{i,u}},{P_{k,2}^{i,u}}),
		\end{align}
		where 
		\begin{align}
		\rho _k^{{\mathop{\rm p}\nolimits}}(z) &= {e_k}(z) + c(z), \\
		\nonumber
		{e_k}(z) &= \sum\limits_{i = 1}^{J_{k|k - 1}^u } {\eta _{k|k - 1}^{i,u}{\textstyle{{s_{k|k - 1}^{i,u}} \over {s_{k|k - 1}^{i,u} + t_{k|k - 1}^{i,u}}}}q_{k,2}(z)}, \\
		\nonumber
		q_{k,2}(z) &= {{\cal N}}(z;{H_k}m_{k|k - 1}^{i,u},{H_k}P_{k|k - 1}^{i,u}H_k^{\top} + {R_k}), \\
		\nonumber
		m_{k,2}^{i,u } &= m_{k|k - 1}^{i,u } + K(z - {H_k}m_{k|k - 1}^{i,u }),\\
		\nonumber
		P_{k,2}^{i,u } &= (I - K{H_k})P_{k|k - 1}^{i,u },\\
		\nonumber
		K &= P_{k|k - 1}^{i,u }H_k^{\top}{({H_k}P_{k|k - 1}^{i,u }H_k^{\top} + {R_k})^{ - 1}},\\
		\nonumber	
		s_{k,2}^{i,u} &= s_{k|k - 1}^{i,u} + 1, \\
		\nonumber
		t_{k,2}^{i,u} &= t_{k|k-1}^{i,u}.
		\end{align}
		\item Misdetection for potentially detected objects:
		\begin{align}
		\omega_k^{j,i}(\emptyset ) &= \omega_{k|k - 1}^{j,i}(1 - r_{k|k - 1}^{j,i} + r_{k|k - 1}^{j,i}{\textstyle{{t_{k|k - 1}^{j,i}} \over {s_{k|k - 1}^{j,i} + t_{k|k - 1}^{j,i}}}}),\\
		r_{k}^{j,i}(\emptyset ) &=  \frac{{r_{k|k - 1}^{j,i}{\textstyle{{t_{k|k - 1}^{j,i}} \over {s_{k|k - 1}^{j,i} + t_{k|k - 1}^{j,i}}}}}}{{1 - r_{k|k - 1}^{j,i} + r_{k|k - 1}^{j,i}{\textstyle{{t_{k|k - 1}^{j,i}} \over {s_{k|k - 1}^{j,i} + t_{k|k - 1}^{j,i}}}}}}, \\
		p_{k}^{j,i}(\emptyset,a) &= \beta (a ;s_{k,3}^{j,i},t_{k,3}^{j,i}){{\cal N}}(x;m_{k,3}^{j,i},P_{k,3}^{j,i}),
		\end{align}
		where
		\begin{align}
		\nonumber
		s_{k,3}^{j,i} &= s_{k|k-1}^{j,i}, \\
		\nonumber
		t_{k,3}^{j,i} &= t_{k|k-1}^{j,i} + 1, \\
		\nonumber
		m_{k,3}^{j,i} &= m_{k|k-1}^{j,i}, \\
		\nonumber
		P_{k,3}^{j,i} &= P_{k|k-1}^{j,i}.
		\end{align}
		\item Update for previous potentially detected objects using received observation set:
		\begin{align}
		\omega_{k}^{j,i}(z) &=\omega_{k|k - 1}^{j,i}r_{k|k - 1}^{j,i}{\textstyle{{s_{k|k - 1}^{j,i}} \over {s_{k|k - 1}^{j,i} + t_{k|k - 1}^{j,i}}}}q_{k,4}(z), \\
		r_{k}^{j,i}(z) &= 1, \\
		p_{k}^{j,i}(x,a|z) &= \beta (a;s_{k,4}^{j,i},t_{k,4}^{j,i}){{\cal N}}(x;m_{k,4}^{j,i},P_{k,4}^{j,i}) ,
		\end{align}
		where
		\begin{align}
		\nonumber
		q_{k,4}(z) &= {{\cal N}}(z;{H_k}m_{k|k - 1}^{j,i},{H_k}P_{k|k - 1}^{j,i}H_k^{\top} + {R_k}), \\
		\nonumber
		m_{k,4}^{j,i} &= m_{k|k - 1}^{j,i} + K(z - {H_k}m_{k|k - 1}^{j,i}), \\
		\nonumber
		P_{k,4}^{j,i} &= (I - K{H_k})P_{k|k - 1}^{j,i}, \\
		\nonumber
		K &= P_{k|k - 1}^{j,i}H_k^{\top}{({H_k}P_{k|k - 1}^{j,i}H_k^{\top} + {R_k})^{ - 1}},\\
		\nonumber
		s_{k,4}^{j,i} &= s_{k|k - 1}^{j,i} + 1,\\
		\nonumber
		t_{k,4}^{j,i} &= t_{k|k - 1}^{j,i}.
		\end{align}
	\end{enumerate}
\end{Pro}
\begin{Rem}
Some properties of the Beta distribution are used during the derivation of update process, e.g., (\ref{pro1}) is used in both parts (a) and (c), and (\ref{pro2}) is used in both parts (b) and (d), in addition, (\ref{pro3}) and (\ref{pro4}) are used in parts (c) and (d) respectively.
\end{Rem}
In terms of the update for potential objects for the first time, to make the form of the Bernoulli component consistent, we approximate the Beta-Gaussian mixture to a single Beta-Gaussian form by performing moment matching as follows.
\begin{align}
p_k^{{\mathop{\rm p}\nolimits}}(x,a|z) = \beta (a;s_{k,2}^{j,i},t_{k,2}^{j,i}){{\cal N}}(x;m_{k,2}^{j,i},P_{k,2}^{j,i}).
\end{align} 
Hereafter, the global hypotheses are constructed based on the obtained single-object hypotheses. As a consequence, the single-object densities from the $i$-th target of the $j$-th global hypothesis at time $k$ are given by
\begin{align}
p_k^{j,i}(x,a)=\beta(a,s_k^{j,i},t_k^{j,i}){\cal N}(x;m_k^{j,i},P_k^{j,i}).
\end{align} 

Further, after each update is finished, component merging is performed by using the Hellinger distance for the Poisson process, meanwhile, component pruning is performed by a predetermined threshold for both Poisson process and MB process. The detailed approximation technology can be found in \cite{Mahler2011}.

In the process of state estimates, the global hypothesis of the MBM process with the highest weight is seletced
\begin{align}
\tilde j = \arg \mathop {\max }\limits_j \prod\limits_i {\omega_k^{j,i}}.
\end{align}
Then, those Bernoulli components whose weights are above a pre-set threshold $\Gamma$,
\begin{align}
\tilde i = \{ i:r_k^{\tilde j,i} > \Gamma \}
\end{align}
are selected as the estimated state of objects, and the estimate for the number of objects is ${{\tilde N}_k} = \sum\nolimits_{\tilde i} {r_k^{\tilde j,\tilde i}}$.

Moreover, the estimate of detection probability can be extracted from the mean of Beta distributions of the selected Bernoulli components,
\begin{align}
\tilde a = \frac{1}{{|{{\tilde N}_k}|}}\sum\nolimits_{\tilde i} {\frac{{s_k^{\tilde j,\tilde i}}}{{s_k^{\tilde j,\tilde i} + t_k^{\tilde j,\tilde i}}}}.
\end{align}
\section{Performance assessment}

In this section, we test the proposed R-PMBM filter and compare it with the R-CPHD filter \cite{Mahler2011} in terms of the Optimal SubPattern Assignment (OSPA) error \cite{Schumacher} with $c=100m$ and $p=1$.

Consider a two-dimensional scenario space $4500m\times4500m$ in which twelve objects move at the nearly constant velocity (NCV) model in the surveillance area. Each object state consists of 2-dimensional position and velocity, i.e., $x = [p_x,p_y,v_x,v_y]^{\top}$ and each  observation polluted by noise is a vector of planar position $z=[z_x,z_y]^{\top}$. Moreover, the parameters of the model (including the dynamic model and observation model) are given by
\begin{align}
\nonumber
{F_k} &= \left[ {\begin{array}{*{20}{c}}
	{{I_2}}&{\Delta {I_2}}\\
	{{0_2}}&{{I_2}}
	\end{array}} \right],Q_k = \sigma _v^2\left[ {\begin{array}{*{20}{c}}
	{{\textstyle{{{\Delta ^4}} \over 4}}{I_2}}&{{\textstyle{{{\Delta ^3}} \over 2}}{I_2}}\\
	{{\textstyle{{{\Delta ^3}} \over 2}}{I_2}}&{{\Delta ^2}{I_2}}
	\end{array}} \right],\\
\nonumber
{H_k} &= \left[ {\begin{array}{*{20}{c}}
	{{I_2}}&{{0_2}}
	\end{array}} \right],{R_k} = \sigma _\varepsilon ^2{I_2},
\end{align}
where $I_n$ and $0_n$ denote the $n\times n$ identity and zero matrices respectively. $\sigma_v=5ms^{-2}$ and $\sigma_\varepsilon =10m$ are the standard deviations of process noise and observation noise. The sampling rate is $\Delta=1s$.  The probability of survival for each object is $p_{S,k} = 0.97$. Besides, the monitored  time of the surveillance area is $T=80s$.

The threshold of Poisson component pruning is $T_P=10^{-5}$ and that of Bernoulli component pruning is $T_B =10^{-5}$. The parameter setting of the R-CPHD filter is the same as those in \cite{Mahler2011}. Moreover, the threshold when extracting object state is set to $\Gamma=0.55$.

The birth model is a Beta-Gaussian mixtures form with eleven Beta-Gaussian components
\begin{align}
{\gamma _k}(x,a) = \sum\limits_{i = 1}^{11} {\eta _b\beta (a;s_b,t_b){{\cal N}}(x;m_{\gamma,k}^i,P_b)}.
\end{align}
All Beta-Gassuain components share the same probability of existence of $\eta_b = 0.03$ and same parameters of Beta distribution of $s_b=t_b=1$, but have the different Gaussian densities. All the Gaussian components have the same covariance matrix of ${P_b} = \text{diag}{({[60,60,60,60]^{\top}})^2}$ but different means, $m_{\gamma,k}^{(1)}=[1000,2300,0,0]^{\top}$, $m_{\gamma,k}^{(2)}=[3000,1200,0,0]^{\top}$,\\ $m_{\gamma,k}^{(3)}=[2000,2000,0,0]^{\top}$, $m_{\gamma,k}^{(4)}=[2000,3500,0,0]^{\top}$,\\ $m_{\gamma,k}^{(5)}=[800,3000,0,0]^{\top}$, $m_{\gamma,k}^{(6)}=[2500,1500,0,0]^{\top}$,\\ $m_{\gamma,k}^{(7)}=[3800,2000,0,0]^{\top}$, $m_{\gamma,k}^{(8)}=[3800,3400,0,0]^{\top}$,\\ $m_{\gamma,k}^{(9)}=[4000,2500,0,0]^{\top}$, $m_{\gamma,k}^{(10)}=[3900,1500,0,0]^{\top}$,\\ $m_{\gamma,k}^{(11)}=[1200,1200,0,0]^{\top}$. 

\begin{figure}[htb]
	\centering
	\includegraphics[width=3.0in]{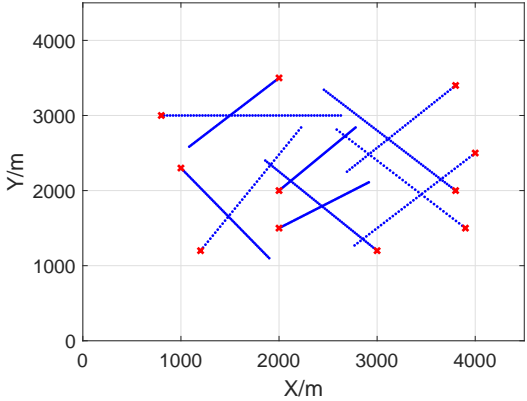}
	\caption{The observed region containing eleven objects, where  $``\times"$ denotes the start point of the trajectory.}
	\label{fig:track}
\end{figure}

Furthermore, clutter is modeled as a Poisson RFS with clutter rate $\lambda_c=10$, which means there are 10 points per scan. Object-originated observations are generated according to a constant detection probability $p_{D}$, which is unknown in all simulation experiments. The observation region and trajectories are presented in Fig. \ref{fig:track}.

\begin{figure*}[!hbp]
	\centering
	\begin{minipage}[b]{0.31\linewidth}
		\centerline{\includegraphics[width=1.08\columnwidth,draft=false]{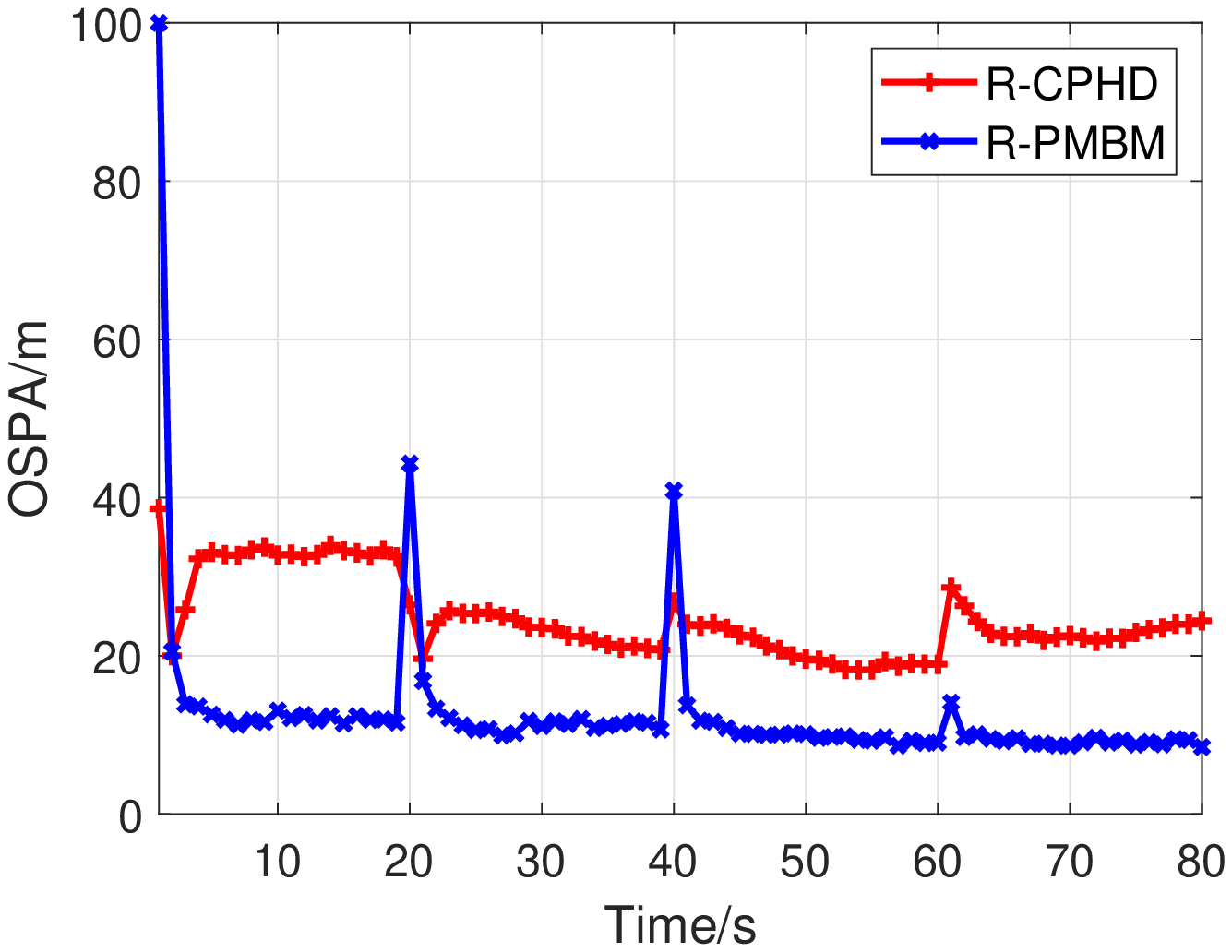}}
		\centerline{(a)}\medskip
	\end{minipage}
	\hfill
	\begin{minipage}[b]{0.31\linewidth}
		\centering
		\centerline{\includegraphics[width=1.08\columnwidth,draft=false]{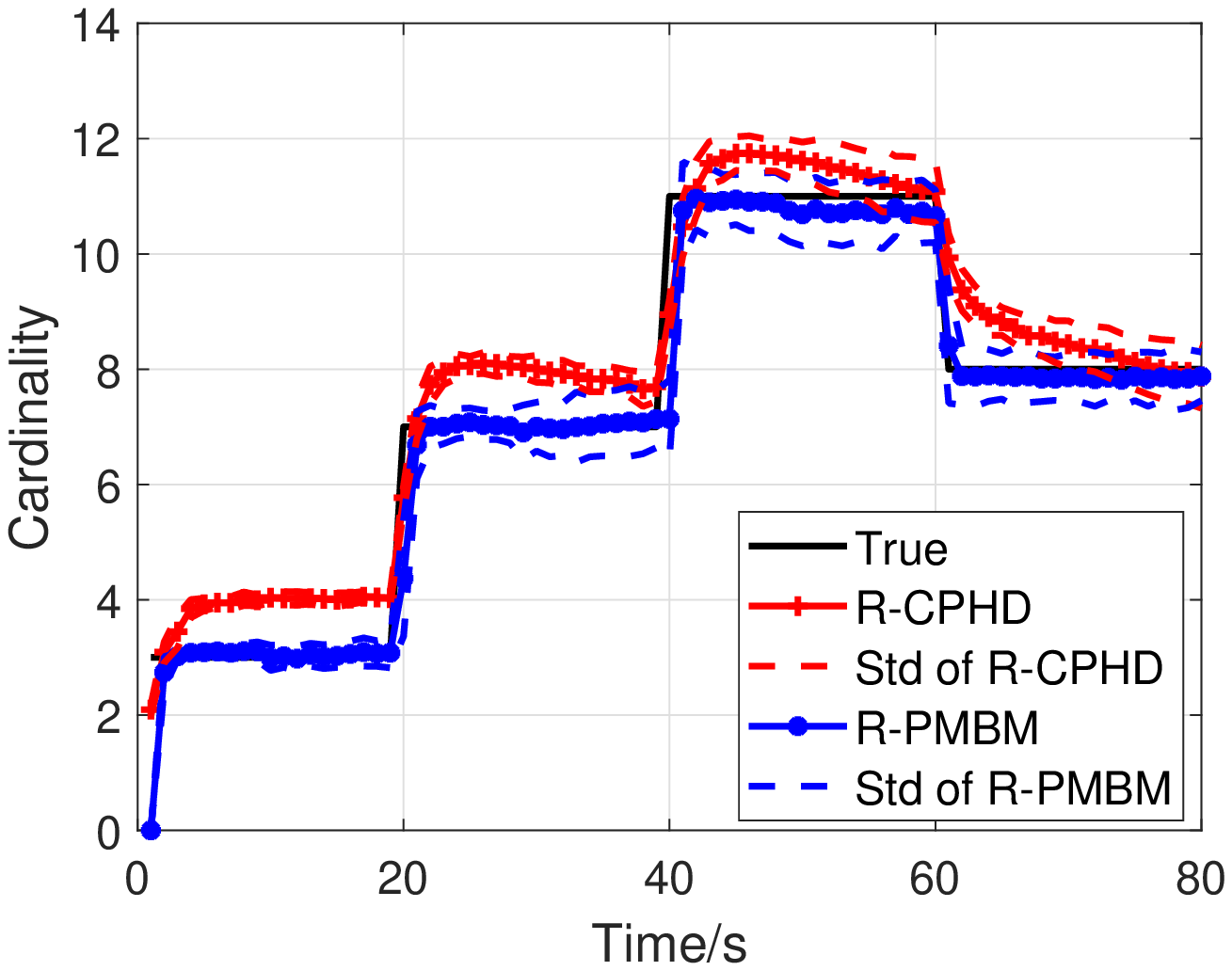}}
		\centerline{(b) }\medskip
	\end{minipage}
	\hfill
	\begin{minipage}[b]{0.31\linewidth}
		\centering
		\centerline{\includegraphics[width=1.08\columnwidth,draft=false]{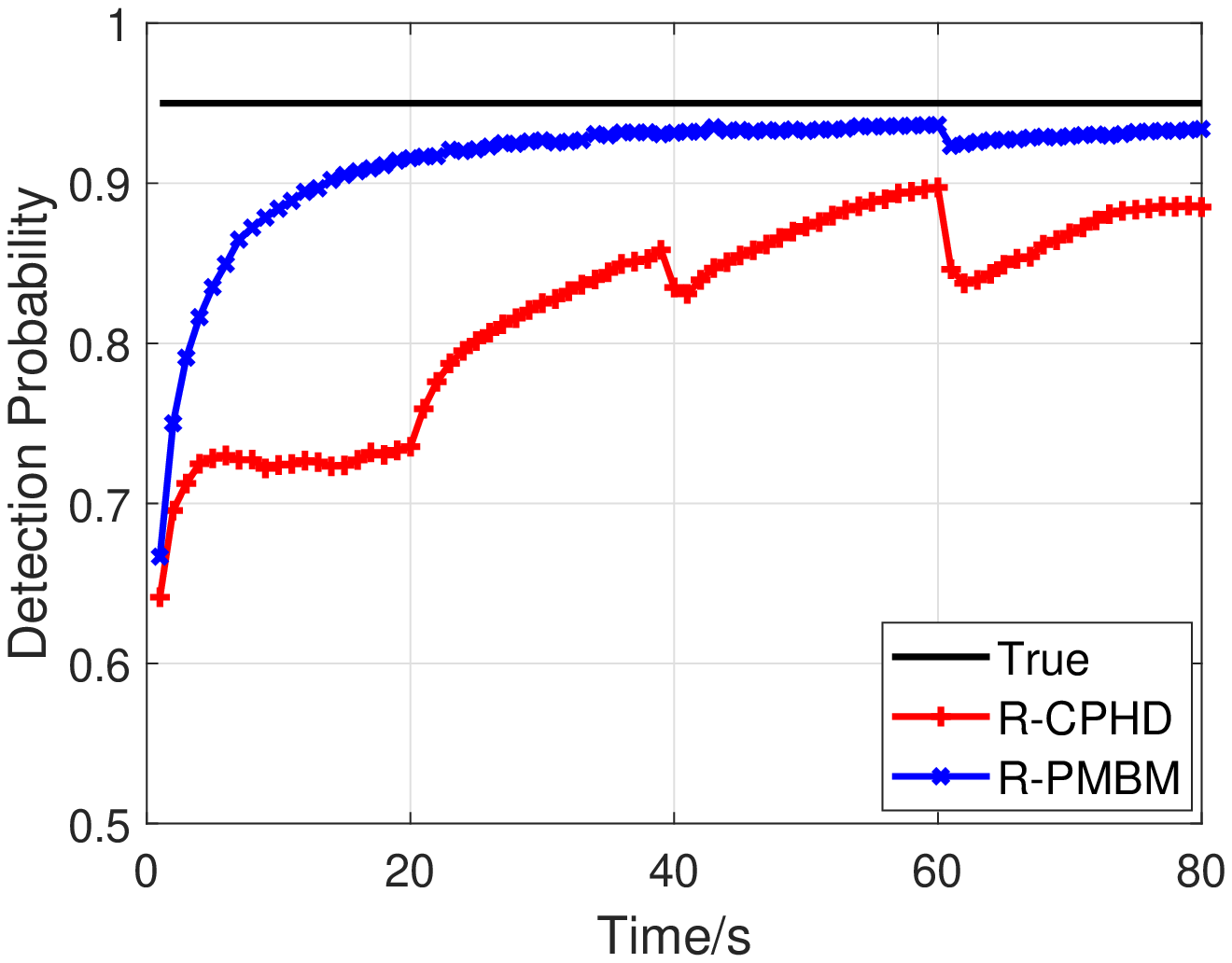}}
		\centerline{(c) }\medskip
	\end{minipage}
	\caption{The comparisons between the R-CPHD and R-PMBM filters with $p_D=0.95$: (a) OSPA errors; (b) cardinality estimate; (c) detection probability.}
	\label{fig:ospa_card_95}
\end{figure*}

Next, two cases with different detection probabilities, $p_D=0.95$ and $p_D=0.65$ respectively, are studied and compared from both OSPA errors and cardinality estimate as well as estimate of detection probability. Besides, the comparisons of different observation noise are also provided. All of the results are averaged over 200 independent Monte Carlo (MC) runs.
\begin{figure}[htb]
	\centering
	\includegraphics[width=3.0in]{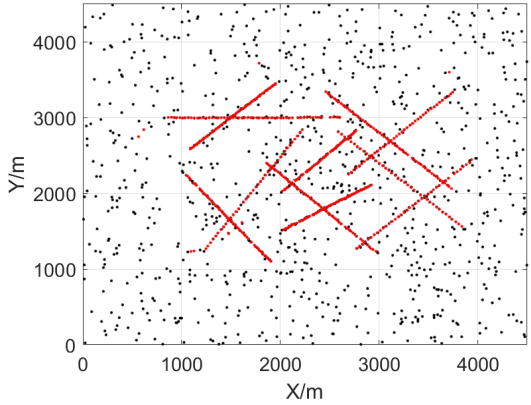}
	\caption{The superposition of the generated observations (\textcolor{black}{black} dots) and estimates of objects (\textcolor{red}{red} dots) in one MC for case 1.}
	\label{fig:track95}
\end{figure}
\subsection{Case 1}
In this scenario, the actual but unknown detection probability is set to $p_D=0.95$. Fig. \ref{fig:track95} shows the superposition of the generated observations during the whole monitored period. The comparisons of OSPA errors and cardinality estimate between the R-CPHD and R-PMBM filters are shown in Fig. \ref{fig:ospa_card_95}. From Fig. \ref{fig:ospa_card_95} (a), the performance of the R-PMBM filter is much better than that of the R-CPHD filter even though the errors are a little relatively large at time $20s$ and $40s$. A possible explanation here is that the R-PMBM filter has a relatively slow response to new objects appear, and so, on average, incurs a higher penalty in this respect. But the R-PMBM filter has lower OSPA errors when the number of objects is steady. Moreover, the comparison of cardinality estimate is shown in Fig. \ref{fig:ospa_card_95} (b), where the overall cardinality estimate of the R-PMBM filter is much better than that of the R-CPHD filter. Furthermore, from Fig. \ref{fig:ospa_card_95} (c), it can be seen that the estimate of $p_D$ for the R-PMBM filter is more precise than the R-CPHD filter and approaches to the true value. In addition, we can also find that the estimate of $p_D$ for the R-PMBM filter suddenly drops when the targets disappear at time $60s$ and then it will converge to the true value again.

Moreover, changing the covariance of observation noise, we compare the OSPA errors given in Table \ref{table1}. It shows that the OSPA errors increase as the covariance of observation noise increases for both filters, which is consistent with expectations. It is worth noting that the performance of the R-PMBM filter is always better than that of the R-CPHD filter under the same parameters.
\begin{table}[!t]
	\begin{center}
		\caption{The comparison of OSPA errors between R-CPHD and R-PMBM filters with $\lambda_c=10$ and different $\sigma_\varepsilon$ for case 1.} \label{table1}
		\begin{tabular}{cccccc}
			\hline
			{\bf\small $\sigma_\varepsilon$ ($p_D=0.95$)} &{5} & {10}  & {15} & {20} & {25} \\
			\hline
			\hline
			{R-CPHD} & 22.02 & 24.76 & 27.81 & 31.62 & 34.36\\
			{R-PMBM} & 8.18 & 12.79 & 15.84 & 17.64 & 21.25\\
			\hline
		\end{tabular}
	\end{center}
\end{table}

In addition, the comparison of different clutter rates is also considered, and the results show given in Table \ref{table2}. Results show that both OSPA errors of two filters increase as $\lambda_c$ increases, meanwhile, the performance of the R-PMBM filter is always better than that of the R-CPHD filter for the same clutte rate.

\begin{table}[!htbp]
	\begin{center}
		\caption{The comparisons of OSPA errors between R-CPHD and R-PMBM filters with $\sigma_\varepsilon =10m$ and different $\lambda_c$ for case 1.} \label{table2}
		\begin{tabular}{cccccc}
			\hline
			{\bf\small $\lambda_c$ ($p_D=0.95$)} &{5} & {10}  & {15} & {20} & {25} \\
			\hline
			\hline
			{R-CPHD} & 30.89 & 33.23 & 33.54 & 34.41 & 34.77\\
            {R-PMBM} & 7.66 & 9.85 & 10.07 & 10.18 & 10.61\\
			\hline
		\end{tabular}
	\end{center}
\end{table}

\begin{figure}[htbp]
	\centering
	\includegraphics[width=3.0in]{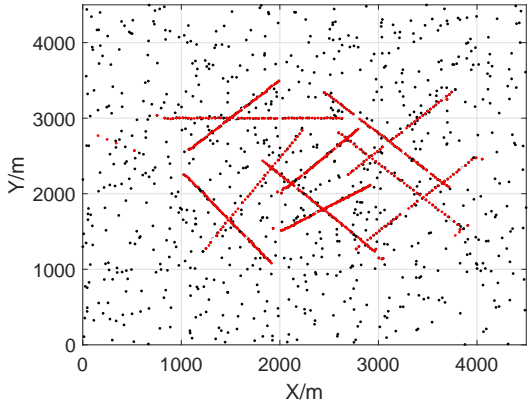}
	\caption{The superposition of the generated observations (\textcolor{black}{black} dots) and estimates of objects (\textcolor{red}{red} dots) in one MC for case 2.}
	\label{fig:track65}
\end{figure}

\subsection{Case 2}
\begin{figure*}[!htbp]
	\centering
	\begin{minipage}[b]{0.31\linewidth}
		\centerline{\includegraphics[width=1.08\columnwidth,draft=false]{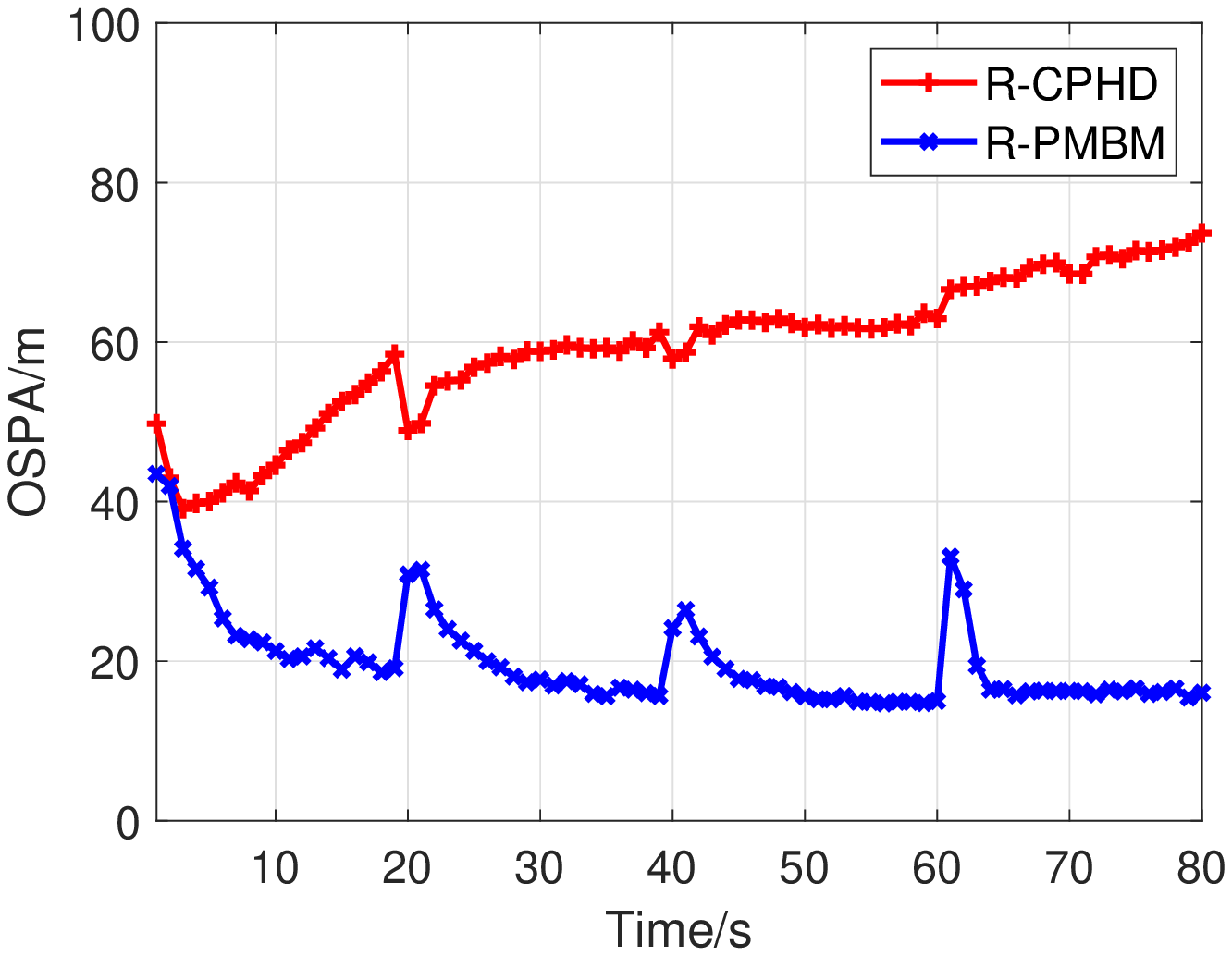}}
		\centerline{(a)}\medskip
	\end{minipage}
	\hfill
	\begin{minipage}[b]{0.31\linewidth}
		\centering
		\centerline{\includegraphics[width=1.08\columnwidth,draft=false]{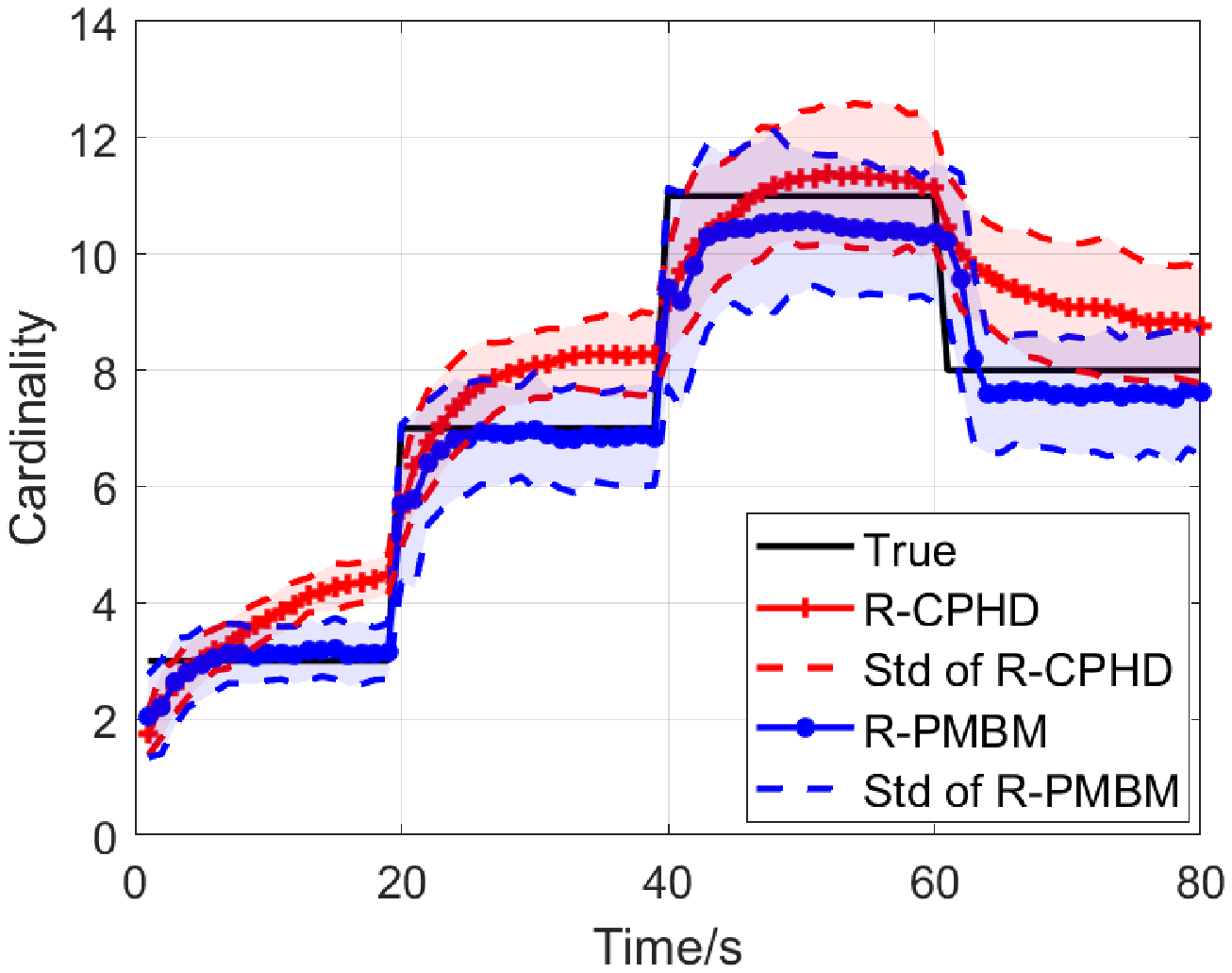}}
		\centerline{(b) }\medskip
	\end{minipage}
	\hfill
	\begin{minipage}[b]{0.31\linewidth}
		\centering
		\centerline{\includegraphics[width=1.08\columnwidth,draft=false]{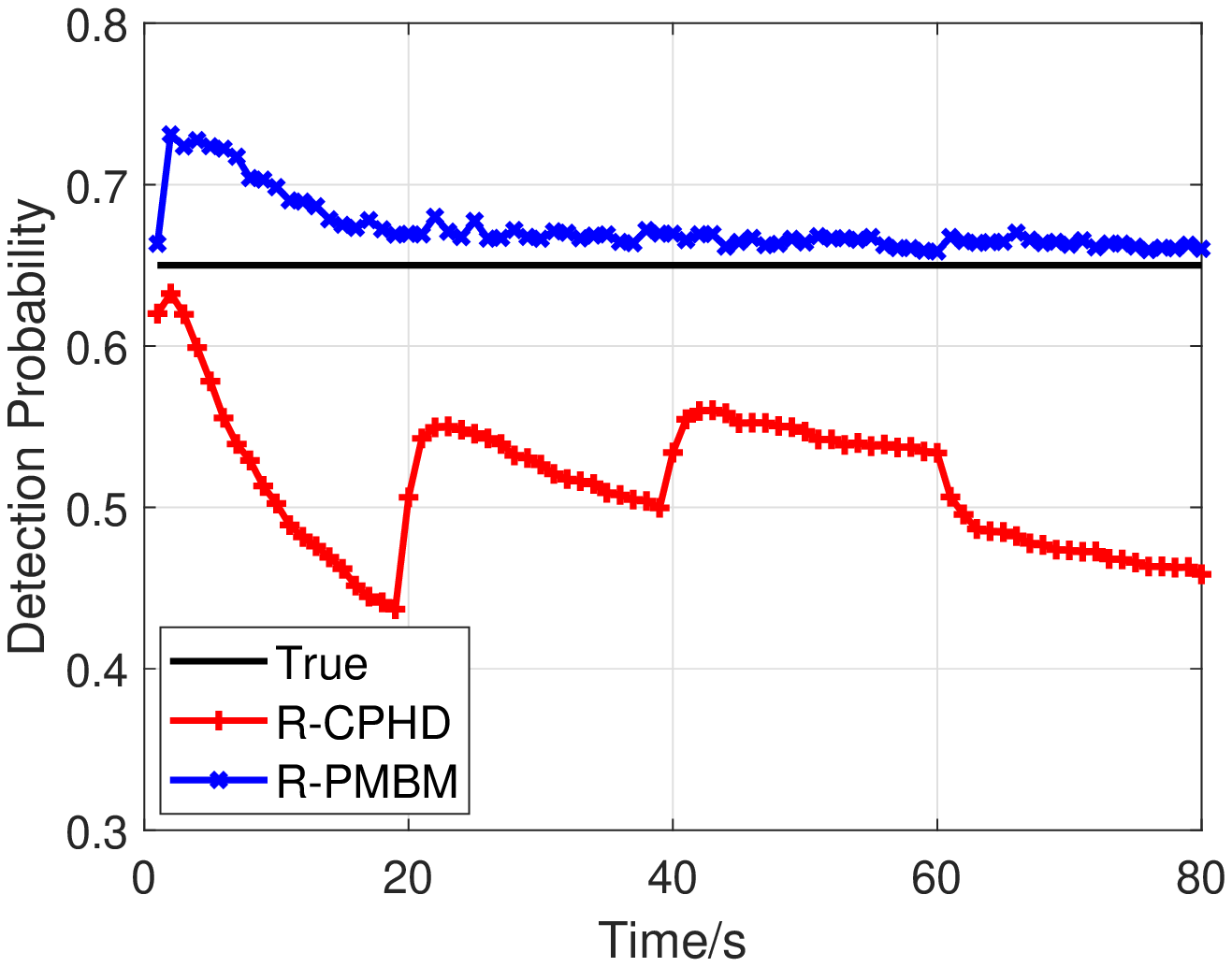}}
		\centerline{(c) }\medskip
	\end{minipage}
	\caption{The comparisons between the R-CPHD and R-PMBM filters with $p_D=0.65$: (a) OSPA errors; (b) cardinality estimate; (c) detection probability.}
	\label{fig:ospa_card_65}
\end{figure*}

Different from case 1, the lower detection probability with $p_D=0.65$ is considered. Fig. \ref{fig:track65} shows the superposition of the observations during the whole monitored period. The comparisons of OSPA errors and cardinality are shown in Fig. \ref{fig:ospa_card_65} (a) and (b). Results show that both OSPA errors and cardinality estimate of the R-PMBM filter are much better than that of the R-CPHD filter, and meanwhile, the gap between two filters is greater compared with case 1. This is because the R-CPHD filter is less able to withstand low detection probability. Besides, the comparison od estimate of $p_D$ is given in Fig. \ref{fig:ospa_card_65} (c), which shows the R-PMBM filter is capable of accurately estimating the $p_D$ whereas the R-CPHD filter almost completely breaks down. The comparisons of averaged OSPA errors under different $\sigma_\varepsilon$ and $\lambda_c$ are also provided in Tables \ref{table3} and \ref{table4}, respectively.
\begin{table}[!htbp]
	\begin{center}
		\caption{The comparisons of OSPA errors between R-CPHD and R-PMBM filters with $\lambda_c=10$ and different $\sigma_\varepsilon$ for case 2} \label{table3}
		\begin{tabular}{cccccc}
			\hline
			{\bf\small $\sigma_\varepsilon$($p_D=0.65$)} &{5} & {10}  & {15} & {20} & {25} \\
			\hline
			\hline
			{R-CPHD} & 57.09 & 59.20 & 60.55 & 62.49 & 64.35\\
			{R-PMBM} & 15.35 & 19.91 & 22.82 & 26.09 & 31.34\\
			\hline
		\end{tabular}
	\end{center}
\end{table}
\begin{table}[!htbp]
	\begin{center}
		\caption{The comparisons of OSPA errors between R-CPHD and R-PMBM filters with $\sigma_\varepsilon=10m$ and different $\lambda_c$ for case 2} \label{table4}
		\begin{tabular}{cccccc}
			\hline
			{\bf\small $\lambda_c$($p_D=0.65$)} &{5} & {10}  & {15} & {20} & {25} \\
			\hline
			\hline
			{R-CPHD} & 58.75 & 59.20 & 59.72 & 59.86 & 60.37\\
			{R-PMBM} & 16.21 & 19.91 & 22.18 & 26.78 & 30.24\\
			\hline
		\end{tabular}
	\end{center}
\end{table}

\section{Conclusions}
In this paper, we mainly research the multi-object tracking (MOT) in unknown detection probability with the Poisson multi-Bernoulli mixture (PMBM) filter. Firstly, a construction strategy by augmenting the state of object with a parameter of detection probability is presented. Then, the recursive expressions are provided including prediction and update processes. Moreover, the detailed implementation by resorting to the Beta-Gaussian mixture technology, where the Beta distribution is used to represent the detection model resulting in Poisson intensity and Bernoulli density functions to be characterized by Beta-Gaussian mixtures and a single Beta-Gaussian form respectively, is provided. Two cases with different detection probabilities are provided to verify the effectiveness of the proposed R-PMBM filter, and also the robustness (covariance of observation noise and clutter rate) has been verified by means of simulation experiments. 
For the future work, the verification of real data for the proposed method may be a worthwhile topic. 


\end{document}